\documentstyle[epsf,aps,twocolumn]{revtex}
\begin{document}
\draft
\preprint{HEP/123-qed}

%%%%%%%%%%%%%%%%%%%%%%%%%%%%%%%%%%%%%%%%%%%%%%%%%%%%%
\twocolumn[\hsize\textwidth\columnwidth\hsize\csname
@twocolumnfalse\endcsname
%%%%%%%%%%%%%%%%%%%%%%%%%%%%%%%%%%%%%%%%%%%%%%%%%%%%%

\title{
Universal Heat Transport in Sr$_2$RuO$_4$
}

\author{
M. Suzuki$^1$, M. A. Tanatar$^{1,2}$\cite{ikhp}, N. Kikugawa$^{1,2}$, Z. Q. Mao$^{1,2}$, Y. Maeno$^{1,2,3}$\cite{email}, and T. Ishiguro$^{1,2}$
}
\address{
$^1$ Department of Physics, Kyoto University, Kyoto 606-8502, Japan \\
$^2$ CREST, Japan Science and Technology Corporation, Kawaguchi, Saitama 332-0012, Japan \\
$^3$ International Innovation Center, Kyoto University, Kyoto 606-8501, Japan \\
}
\date{\today}
\maketitle

\begin{abstract}

We present the temperature dependence of the thermal conductivity $\kappa(T)$ of the unconventional superconductor Sr$_2$RuO$_4$ down to low temperatures ($\sim$100 mK). In $T \rightarrow 0$ K limit we found a finite residual term in $\kappa/T$, providing clear evidence for the superconducting state with an unconventional pairing.  The residual term remains unchanged for samples with different $T_{\rm c}$, demonstrating the universal character of heat transport in this spin-triplet superconductor.  The low-temperature behavior of $\kappa$ suggests the strong impurity scattering with a phase shift close to $\pi /2$.  A criterion for the observation of universality is experimentally deduced.  

\end{abstract}


%%%%%%%%%%%%%%
]\narrowtext
%%%%%%%%%%%%%%

Through the study of the cuprate superconductor in the last decade, a notable progress has been made in an understanding of the thermodynamic and transport properties of quasi-particles (QP) in unconventional superconductors.  One of the important findings is the novel phenomenon called the universal transport, pointed out first by Lee \cite{Lee}.  In an unconventional superconductor with nodes in the superconducting (SC) gap, non-magnetic impurities suppress the transition temperature ($T_{\rm c}$) and induce the finite density of QP states at the Fermi level with the energy width $\gamma_{\rm imp}$ \cite{Balatsky}.  The presence of impurities increases both the QP density and the impurity scattering rate, which completely cancel each other \cite{Sun,Graf}. In this case the QP transport, such as thermal conductivity ($\kappa$), restores the temperature ($T$) dependence typical for the normal state (e.g. $\kappa \propto T$) \cite{Gottwick} and becomes independent of impurity concentration.  The magnitude of the residual term in $\kappa/T$ depends only on the QP spectrum near the gap nodes, providing a useful tool for studying the SC order parameter\cite{Graf}. 

Experimentally, the universal residual term in $\kappa/T$ at $T \rightarrow$ 0 K has been reported in optimally-doped high-$T_{\rm c}$ cuprates YBa$_2$Cu$_3$O$_{6.9}$ \cite{Taillefer} and Bi$_2$Sr$_2$CaCu$_2$O$_8$ \cite{BehniaBi,Nakamae}.  Their residual values are consistent with the $d$-wave SC gap structure \cite{Chiao}.  For non-$d$-wave gap symmetries, the universal conductivity has never been observed. In contrast, the extrapolated $\kappa/T$ at $T \rightarrow$ 0 K of the spin-triplet superconductor UPt$_3$ \cite{SuderowLT97,SuderowLT99} rapidly increases with the density of defects, showing no universal behavior, and the residual term seems to vanish in zero-disorder limit \cite{SuderowLT99}. The origin of the lack of universality is still unclear, similar to the case of under-doped cuprates \cite{Hussey}, and it is strongly desirable to extend studies to other unconventional superconductors. 

In this respect, of special interest is the study of Sr$_2$RuO$_4$, a unique quasi-two-dimensional (Q2D) spin triplet superconductor \cite{IshidaNature,Duffy}. This material has a layered perovskite structure \cite{MaenoNature} and its electronic state is well described in the Landau-Fermi liquid model \cite{MackenzieFermi}. The SC state of Sr$_2$RuO$_4$ ($T_{\rm c}$ = 1.5 K) is chiral with a broken time reversal symmetry \cite{Luke} and its $T_{\rm c}$ is rapidly suppressed by non-magnetic impurities \cite{Mackenzie,MaoDefect}. Views on the SC gap symmetry of Sr$_2$RuO$_4$ changed in time. Early theoretical prediction of the fully gapped $p$-wave chiral state \cite{Rice} did not find support in recent experiments, suggesting the SC gap with line nodes \cite{Zaki,Ishida,Bonalde,Tanatar}, likely running parallel to the conducting plane \cite{TanatarPRL,Izawa,Lupien}. Several new theoretical models have been proposed \cite{Agterberg,Miyake,Hasegawa,Zhitomirsky} to explain the experimentally observed "triplet-chiral-nodal" SC order parameter. 

The thermal conductivity measurement of Sr$_2$RuO$_4$ down to 20 mK was reported by Suderow {\it et al}. \cite{SuderowSRO}. This study was restricted to strongly disordered samples with $T_{\rm c} <$ 1 K.  Measurements with high quality samples, with $T_{\rm c} >$ 1.4 K, have been performed only down to 0.3 K \cite{Tanatar,TanatarPRL,Izawa}, which is insufficient for the estimation of the residual term.  

In this Letter we report the systematic study of the in-plane thermal conductivity of Sr$_2$RuO$_4$ down to 100 mK on samples with widely varying quality, with $ T_{\rm c}$ from 0.7 K to 1.5 K.  We found a universal behavior of the thermal conductivity in this material with non-$d$-wave order parameter. The finite and universal $\kappa/T$ at $T \rightarrow$ 0 K in the zero impurity limit provides a key support for the SC gap with line nodes. Negligibly small phonon contribution to the total $\kappa$ allowed us to characterize the $T$ dependence of the electronic $\kappa$, which was not possible up to now in the high-$T_{\rm c}$ cuprates  \cite{Nakamae}. We found a reasonable agreement with the theoretical predictions by Graf {\it et al.} \cite{Graf} in the unitarity scattering limit.

Single crystals of Sr$_2$RuO$_4$ were grown 
in an infrared image furnace by the floating zone method \cite{MaoCrystal}.  
We used samples of 5 different batches with $T_{\rm c}$ = 1.44 K, 1.32 K, 1.27 K, 1.09 K and 0.71 K, denoted as \# 1 - 5, respectively. The $T_{\rm c}$ and the transition width $\delta T_{\rm c}$ were defined at the peak temperature and the full width at half maximum of the imaginary part of the AC susceptibility, respectively (Table I). The difference in $T_{\rm c}$ comes from the variation of the density of non-magnetic impurities, such as Al or Si \cite{Mackenzie}, and/or crystalline defects \cite {MaoDefect}. The systematic studies on the electrical resistivity $\rho$ and resistive $T_{\rm c}$ revealed that the suppression of $T_{\rm c}$ is well described by the Abrikosov-Gorkov type equation \cite{Hirschfeld}, 
\begin{eqnarray}
\ln(\frac{T_{\rm c0}}{T_{\rm c}}) = \Psi(\frac{1}{2}+\frac{\hbar{\it \Gamma}}{2\pi k_{\rm B}T_{\rm c}})-\Psi(\frac{1}{2}), \label{AG}
\end{eqnarray}
where $\Psi(x)$ is the digamma function, $T_{\rm c0}$ = 1.5 K is maximum $T_{\rm c}$ for the disorder free material and ${\it \Gamma}$ is the non-magnetic impurity scattering rate in the normal state.  The values of ${\it \Gamma}$ for the samples under study, deduced from Eq. (\ref{AG}), are listed in Table I.  

The thermal conductivity was measured along the in-plane [100] direction between 100 mK and 2 K by the one-heater and two-thermometer method described elsewhere \cite{Tanatar}. Typical dimensions of the samples were 2.0 mm$\times$0.5 mm$\times$0.08 mm, with the longest direction corresponding to the [100] crystal axis and the shortest to the [001] axis. 

Fig. \ref{Td} shows the $T$ dependence of $\kappa/T$ of samples \# 1 to 5, with the $T_{\rm c}$ indicated by arrows. All the samples show almost flat $T$ dependence of $\kappa/T$ in the normal state, as expected for the impurity scattering regime, except for sample \# 1 showing slight decrease with $T$ indicative of small electron-electron scattering. The normal state conductivity increases with $T_{\rm c}$, i.e. with decreasing ${\it \Gamma}$. Below $T_{\rm c}$, $\kappa/T$ decreases monotonically and slowly. In sample \# 1, $\kappa/T$ at 0.3 K (0.2 $T_{\rm c}$) is $\sim$ 5 W/K$^2$m (0.2 $\kappa(T_{\rm c})/T_{\rm c}$), which should be practically zero for the fully gapped superconductor \cite{Satterthwaite}. Although the normal state $\kappa/T$ decreases 4 times with increasing ${\it \Gamma}$, in the SC state all $\kappa/T$ curves converge to a finite value at around 2 $\pm$ 0.5 W/K$^2$m at very low $T$.  

The $\rho$ and $\kappa/T$ measurements made on the same contacts in zero $H$ above $T_{\rm c}$ revealed that the Wiedemann-Franz ratio $\rho\kappa/T$ at $T_{\rm c}$ is satisfied within accuracy of our determination ($\pm10 \%$ \cite{accuracy}) for all sample qualities under study, showing a dominance of the electronic transport in the normal state at temperatures as high as $T_c$. In the SC state, the phonon conductivity $\kappa^{\rm g}$ can be estimated in the boundary scattering limit, providing an upper bound for $\kappa^{\rm g}$, by using a simple kinetic formula $\kappa^{\rm g}_{\rm B} = \frac{1}{3}C_{\rm ph} \bar{v} l $, where $C_{\rm ph}$ is the phonon specific heat per unit volume proportional to $T^3$, $\bar{v}$ is the averaged sound velocity and $l$ is the phonon mean free path, which is of the order of the sample dimension. Since both $\bar{v}$ and $l$ are temperature independent, it can be expressed as $\kappa^{\rm g}_{\rm B}/T = \lambda^{\rm g}_2 T^2$ with a constant coefficient $\lambda^{\rm g}_2$.  Taking the phonon specific heat coefficient of 0.197 mJ/K$^4$mole \cite{Zaki} and $\bar{v}$=1400 m/s \cite{Matsui,Carruthers}, we obtain $\lambda^{\rm g}_2 = 0.2\ {\rm W/K^4m}$, which is a small fraction of the total $\kappa$ in all samples.  

On $T$ increase, $\kappa/T$ vs. $T$ deviates from the universal value, shows an upward curvature up to 0.3 K and then turns into a linear increase at higher $T$. The $\kappa/T$ vs. $T^2$ dependence (Fig. \ref{Td} inset) shows that super-linear portion of the $\kappa/T$ vs. $T$ curve is well reproduced by a polynomial expression $\kappa/T = \lambda^{\rm e}_0 + \lambda^{\rm e}_2 T^2$ (we show only the data for samples \# 1 and \# 3 for clarity).  We should notice here that, despite having the same $T^2$ dependence as expected for $\kappa^{\rm g}/T$, the observed $\kappa/T$ is completely electronic in origin, which is of special interest in this study. Actually measured $\lambda^{\rm e}_2$ decreases with decreasing $T_{\rm c}$ from 44 W/K$^4$m for sample \# 1 to 8 W/K$^4$m for sample \# 5, but still is 40 times larger than $\lambda^{\rm g}_2$ of 0.2 W/K$^4$m.  The finite residual coefficient $\lambda^{\rm e}_0$ indicates existence of QP at $T$=0 and, consequently, the non-$s$-wave pairing. In discussion of the residual term which follows below, we use a conventional notation, $\lambda^{\rm e}_0 \equiv \kappa_{00}/T$.  

In Fig. \ref{Universal} we plot (filled squares) $\kappa_{00}/T$ as a function of the impurity scattering rate divided by the impurity-free $T_{\rm c}$, $\hbar{\it \Gamma}/k_{\rm B}T_{\rm c0}$.  The vertical error bars reflect the scatter of observed $\kappa/T$ and the horizontal error comes from the transition width $\delta T_{\rm c}$.  For the reference, we show with filled triangles $\kappa/T$ by Suderow {\it et al}. \cite{SuderowSRO} extrapolated at $T \rightarrow$ 0 K.  It can be clearly seen that {\it the residual thermal conductivity $\kappa_{00}/T$ remains unchanged for $\hbar{\it \Gamma}/k_{\rm B}T_{\rm c0}$} $<$ 0.2.  Moreover, $\kappa_{00}/T$ values converge to a finite value of 1.7 $\pm 0.15$ W/K$^2$m in the $\hbar{\it \Gamma}/k_{\rm B}T_{\rm c0} \rightarrow 0$ limit, giving strong evidence for the {\it nodal} gap structure. The finite $\kappa_{00}/T$ should appear even in the superconductors with fully gapped non-$s$-wave state (e.g., isotropic 2D $p$-wave state \cite{Rice}) when a small number of impurities exist, but it should vanish in the ${\it \Gamma} \rightarrow 0$ limit with breaking of the universality \cite{Maki}.  

Theoretically, the universal residual term $\kappa_{00}/T$ for the nodal superconductor is explicitly expressed as
\begin{eqnarray}
\frac{\kappa_{00}}{T} = \frac{\pi^2k_{\rm B}^2}{3}N_{\rm F}v_{\rm F}^2\frac{a\hbar}{2\mu\Delta_{0}}, \label{residual}
\end{eqnarray}
where $N_{\rm F}$ and $v_{\rm F}$ are the density of states in the normal state and the Fermi velocity, respectively.  $\Delta_0$ is the maximum amplitude of the gap, $\mu$ is the slope of the gap at the node defined as $\mu \equiv \frac{1}{\Delta_0}\frac{\partial \Delta(\phi)}{\partial \phi}$, where $\phi$ is the in-plane angle of the wave vector on the circular Fermi surface, and $a$ is the coefficient of the order of unity which depends on the gap symmetry \cite{Graf}.  Thus from the residual value $\kappa_{00}/T$ we can obtain the quantity related to the gap structure, $\mu\Delta_0/a$.   

By assuming that the SC gaps have the same magnitude on the three Fermi surfaces of Sr$_2$RuO$_4$, and taking experimentally measured Fermi velocities and effective masses for respective bands \cite{MackenzieFermi}, we can reproduce the observed residual value of $\kappa_{00}/T$ = 1.7 W/K$^2$m with $\mu\Delta_{0}/a$ = 0.39 meV.  This value is close to typical values for several nodal gap structures. For the 2D $f$-wave gap with four vertical line nodes $\Delta=\Delta_0\cos(2\phi)$ \cite{Hasegawa}, $a$ is $4/\pi$ \cite{Graf}, $\mu$ = 2, $\Delta_0 = 2.14k_{\rm B}T_{\rm c}$ in the weak coupling limit and we obtain $\mu\Delta_{0}/a$ = 0.43 meV. For the state with horizontal line node, indicated in recent directional experiments \cite{TanatarPRL,Izawa,Lupien}, the universal thermal conductivity was not considered theoretically. However, we can estimate the residual value of $\kappa_{00}/T$ by assuming that in the limit of the small Fermi velocity along the [001] axis $v_{{\rm F} \bot} \rightarrow$ 0, the 3D polar state $\Delta=\Delta_0\cos(\theta)$ ($\theta$ is polar angle measured from the [001] axis) is a good representation of a Q2D horizontal nodal state. By using the parameters for the polar state \cite{Graf} $a$=1, $\mu=1$ and $\Delta_0 = 2.46k_{\rm B}T_{\rm c}$, we obtain $\mu\Delta_{0}/a$ = 0.32 meV, which is also close to the experimentally observed value. Thus we conclude that the observed residual term comes from the contribution around the gap nodes.  

Recently Zhitomirsky and Rice \cite{Zhitomirsky} proposed the orbital dependent gap structure, which is a modified extension of the original theory by Agterberg {\it et al.} \cite{Agterberg}.  They assumed the line nodes only on two of three Fermi surfaces, $\alpha$ and $\beta$. If we consider that only these two sheets make contribution to $\kappa_{00}/T$ = 1.7 W/K$^2$m, we obtain $\mu\Delta_{0}/a$ = 0.27 meV.  They also assumed a very small sub-Kelvin amplitude of the gap $\Delta_{0}$ on these sheets; thus to explain the observed value of $\kappa_{00}/T$ we need a relatively steep slope $\mu$ at the nodes. 

Since the observed $\kappa/T$ is almost purely electronic, we can compare the experimental results with available theories for the electronic thermal conductivity. Two theories predict a $T^2$ term in the electronic $\kappa/T$. The model by Zhitomirsky and Walker \cite{ZhitomirskyWalker}and by Graf and Balatsky \cite{GrafSRO} (ZWGB) is valid in the range of $\gamma_{\rm imp} < k_{\rm B}T$ due to the QP density coming from the nodal SC gap in case of unitary scattering. The model by Graf {\it et al}. \cite{Graf} considers finite $T$ correction to the residual conductivity $\kappa_{00}/T$ in the range of $k_{\rm B}T < \gamma_{\rm imp}$, where the coefficient of the $T^2$ term of $\kappa/T$ strongly depends on the impurity scattering phase shift \cite{Graf}.  As can be seem from Inset in Fig. \ref{Td}, the $T^2$ term is observed from the lowest $T$, directly above universality limit. This implies that the condition $\gamma_{\rm imp} < k_{\rm B}T$ required for observation of the $T^2$ term in ZWGB model is not met. Therefore, we stick in interpretation to the Graf model \cite{Graf}. Here the $T^2$ term is observed as long as we satisfy the condition for $k_{\rm B}T<\gamma_{\rm imp}$, which corresponds to $T < 0.3$ K in our experiment. This relatively large impurity band means that the scattering is strong. Indeed, in the unitarity limit (the phase shift $\delta = \pi/2$) the impurity bandwidth increases rapidly with the impurity scattering rate as $\gamma_{\rm imp} \sim \sqrt{\Delta_0 \hbar {\it \Gamma}}$ \cite{Balatsky}, which corresponds to about 0.5 K for sample \# 1, while in the Born limit ($\delta = 0$) the impurity bandwidth becomes exponentially small, $\gamma_{\rm imp} \sim \Delta_0\exp(-\frac{\Delta_0}{\hbar{\it \Gamma}})$ \cite{Balatsky}.  Experimentally, the impurity bandwidth can be estimated from the parameters of the fitting formula, $\kappa/T = \lambda^{\rm e}_0 + \lambda^{\rm e}_2 T^2$ introduced above.  In the unitarity scattering limit, the deviation of $\kappa/T$ from the universal value at finite $T$ is described as \cite{Graf}
\begin{eqnarray}
\frac{\kappa}{T} = \frac{\kappa_{00}}{T}(1+\frac{7\pi^2}{60}(\frac{k_{\rm B}T}{\gamma_{\rm imp}})^2).		\label{tempdep}
\end{eqnarray}
The $\gamma_{\rm imp}$ deduced by comparing Eq. (\ref{tempdep}) with the coefficients $\lambda^{\rm e}_0$ and $\lambda^{\rm e}_2$ for each sample is plotted in Fig. \ref{Universal} with an open circle.  The uncertainty due to the possible small phonon contribution is at most 10 \%.  The least square fit with the  $\sqrt{{\it \Gamma}}$ function $\gamma_{\rm imp} = 0.7\sqrt{\hbar{\it \Gamma}/k_{\rm B}T_{\rm c0}}$ is drawn with a dotted line.  The rapid increase and $\sqrt{{\it \Gamma}}$ dependence of $\gamma_{\rm imp}$ support the strong impurity scattering and verify the deduction of $\gamma_{\rm imp}$ from Eq. (\ref{tempdep}), while the absolute value is about two times smaller than the theoretical value of $\gamma_{\rm imp} \sim \sqrt{\Delta_0 \hbar{\it \Gamma}}$.  A slight deviation of $\delta$ from $\pi/2$ makes the impurity bandwidth exponentially small at small ${\it \Gamma}$ \cite{Graf} and thus the non-negligible $\gamma_{\rm imp}$ of sample \# 1 ($\hbar{\it \Gamma}/k_{\rm B}T_{\rm c0}$ = 0.051) implies the large scattering phase shift $\delta = \pi/2 (\pm 10 \%)$.  
  
From our data we can derive the explicit condition of the low $T$ and small $\gamma_{\rm imp}$ "universality limit" which is theoretically described as $k_{\rm B}T \le \gamma_{\rm imp} \ll \Delta_0$. In Fig. \ref{Universal} we can see that above $\hbar{\it \Gamma}/k_{\rm B}T_{\rm c0}$ = 0.2 the universality starts to break and this crossover point corresponds to $\gamma_{\rm imp}/k_{\rm B} \sim 0.3\ {\rm K} \sim 0.1\Delta_0/k_{\rm B}$.  Thus $k_{\rm B}T \le \gamma_{\rm imp} \le 0.1\Delta_0$ can be considered as an experimentally deduced criterion for the observation of universal conductivity.  

The deviation of $\kappa_{00}/T$ from the universal value with increasing $\hbar{\it \Gamma}/k_{\rm B}T_{\rm c0}$ is small, even for the largest $\hbar{\it \Gamma}/k_{\rm B}T_{\rm c0}$, giving at most 50~\% increase for sample \# 5 ($\hbar{\it \Gamma}/k_{\rm B}T_{\rm c0}$ = 0.6).  According to the calculation by Sun and Maki \cite{Sun} in the unitarity scattering limit, $\kappa_{00}/T$ increases rapidly with ${\it \Gamma}$ above the universality limit and it reaches almost twice of the universal value at $\hbar{\it \Gamma}/k_{\rm B}T_{\rm c0}$ = 0.6.   Similar slow increase of $\kappa_{00}/T$ was also pointed out in YBa$_2$Cu$_3$O$_{6.9}$ \cite{Taillefer}.  While the slight deviation of the scattering phase shift from $\pi/2$ might be able to explain it \cite{Taillefer}, the calculated steep increase of $\kappa_{00}/T$ \cite{Sun} is inappropriate for the description of Sr$_2$RuO$_4$.  The critical resistivity, above which $T_{\rm c}$ disappears, is about 1 $\mu\Omega$cm \cite{Mackenzie,MaoDefect} and the corresponding critical $\kappa/T$ is obtained by the Wiedemann-Franz law as 2.4 W/K$^2$m, which is almost equal to our largest $\kappa_{00}/T$.  Since there is no clear reason for $\kappa/T$ at $T$=0 K to become higher in the SC state than in the normal state, the increase of $\kappa_{00}/T$ with ${\it \Gamma}$ expected in Ref. \cite{Sun} should be limited. This point should be clarified in the further studies.  

In conclusion, we experimentally observed universal thermal conductivity in Sr$_2$RuO$_4$, indicating that this phenomenon is not restricted to $d$-wave gap symmetries. The universality gives clear evidence for the nodal gap.  From the analysis of the low temperature thermal conductivity within the framework of the theory by Graf {\it et al}. \cite{Graf}, we extract the impurity bandwidth and the scattering phase shift close to $\pi/2$.  We experimentally derive a criterion for the deviation from the universal limit. 

The authors are grateful to M. B. Walker for useful discussions and comments.  They also thank Y. Shimojo for technical supports.  This work was partly supported by the Grant-in-Aid for Scientific Research on Priority Area "Novel Quantum Phenomena in Transition Metal Oxides" from the Ministry of Education, Culture, Sports, Science and Technology Japan.  

%%%%%%%%%%%%%%%%%%%%%%%%%%%%%%%%%%%%%%%%%%%%%%%%

\begin{table}[htbp]
 \caption{The properties of 5 single crystals of Sr$_2$RuO$_4$.  $T_{\rm c}$ and $\delta T_{\rm c}$ were determined from the AC susceptibility measurements.  The impurity scattering rate normalized by the maximum $T_{\rm c}$, $\hbar{\it \Gamma}/k_{\rm B}T_{\rm c0}$ is deduced from Eq. (\ref{AG}). }\label{table}
 \begin{center}
  \begin{tabular}{cccccc}
      & \# 1  & \# 2  & \# 3 & \# 4 &  \# 5 \\
    \hline
     $T_{\rm c}$ (K) & 1.44 & 1.32  & 1.27 & 1.09 & 0.71 \\

     $\delta T_{\rm c}$ (K)  & 0.02 &  0.03  & 0.03  & 0.05 & 0.15  \\

     $\hbar{\it \Gamma}/k_{\rm B}T_{\rm c0}$   & 0.051 & 0.15 &   0.20  & 0.35 & 0.60   
  \end{tabular}
 \end{center}
\end{table}

\begin{figure}
    \begin{center}
    \epsfxsize=7cm
    \epsfbox{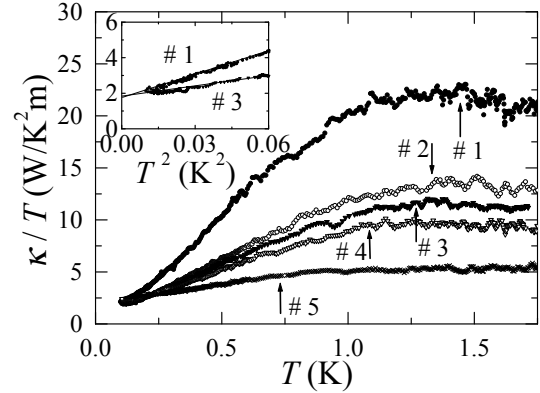}
	\end{center}
\caption{The temperature dependence of the thermal conductivity $\kappa/T$ along the [100] direction for 5 samples of different quality. The arrows denote the transition temperatures $T_{\rm c}$. Inset shows the $\kappa/T$ vs. $T^2$ plot for samples \# 1 and \# 3. }\label{Td}
\end{figure}

\begin{figure}
    \begin{center}
    \epsfxsize=7.5cm
	\epsfbox{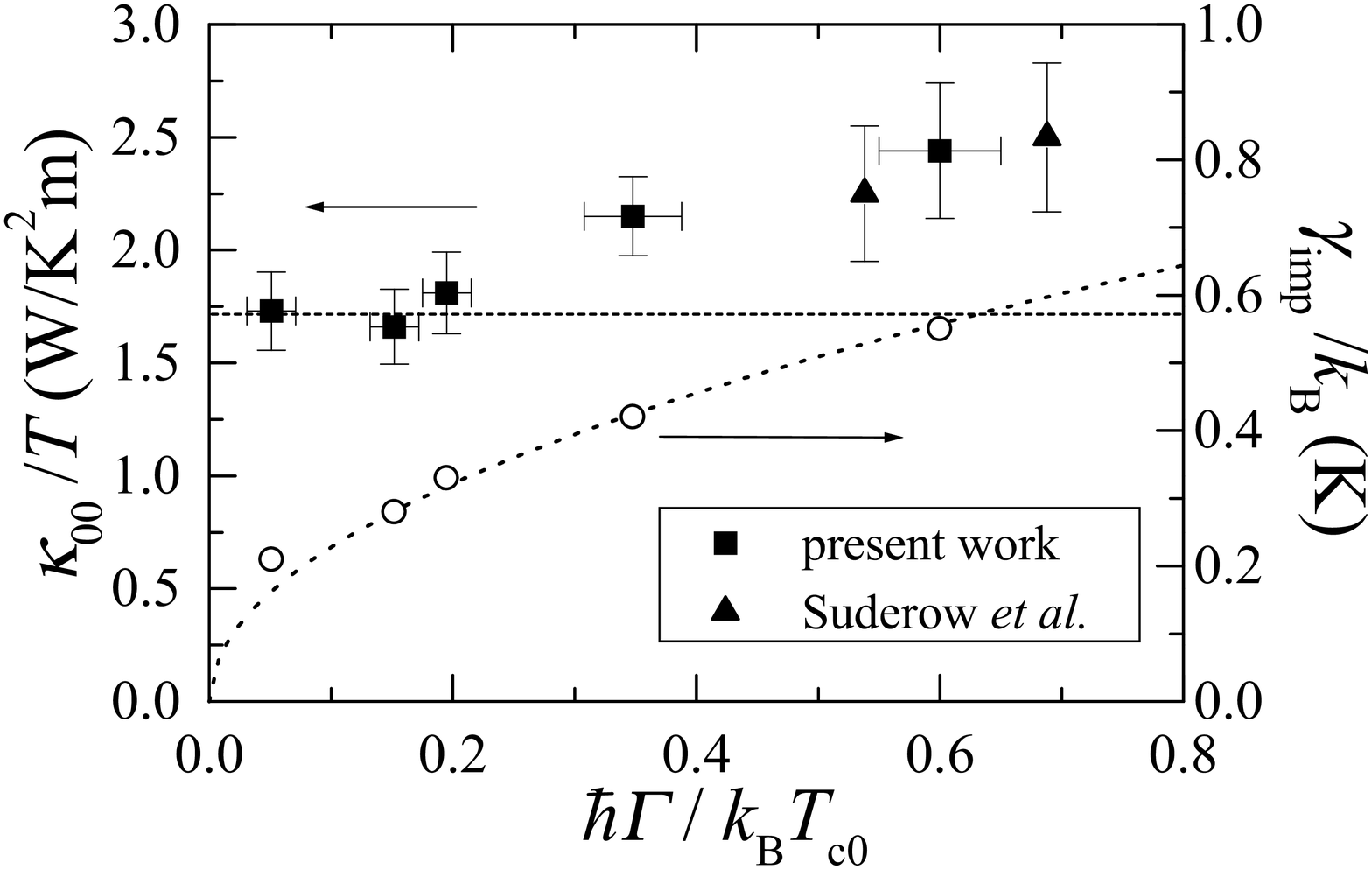}
	\end{center}
\caption{The residual thermal conductivity $\kappa_{00}/T$ and the impurity bandwidth $\gamma_{\rm imp}$ as a function of scattering rate $\hbar{\it \Gamma}/k_{\rm B}T_{\rm c0}$.  Filled squares show $\kappa_{00}/T$ of this study, filled triangles are by Suderow {\it et al} [32]. The open circles represent $\gamma_{\rm imp}$ and the dotted line is a fit with the $\sqrt {\it \Gamma}$ function.  }\label{Universal}
\end{figure}


\begin{thebibliography}{99}
\bibitem{accuracy} The uncertainty is mainly coming from the accuracy of $\rho$ measurements on these low resistance samples.
\end{thebibliography}
\end{document}